\documentclass[a4paper,fleqn,usenatbib]{mnras}

\usepackage{newtxtext,newtxmath}

\usepackage[T1]{fontenc}
\usepackage{ae,aecompl}
\usepackage{lastpage}

\usepackage{graphicx}	
\usepackage{amsmath}	
\usepackage{amssymb}	

\title[Statistical-likelihood Exo-Planetary Habit. Index]{Statistical-likelihood Exo-Planetary Habitability Index (SEPHI)}

\author[J. M. Rodr\'iguez-Mozos and A. Moya]{
J. M. Rodr\'iguez-Mozos,$^{1}$
A. Moya,$^{2}$\thanks{E-mail: atreyu0@gmail.com (AM)}
\\
$^{1}$Universidad Internacional de Valencia (VIU). 46021. Valencia. Spain\\
$^{2}$University of Granada (UGR). Dept. Theoretical Physics and Cosmology. 18071. Granada. Spain
}

\date{Accepted XXX. Received YYY; in original form ZZZ}

\pubyear{2017}

\begin{document}
\label{firstpage}
\pagerange{\pageref{firstpage}--\pageref{lastpage}}
\maketitle

\begin{abstract}
A new Statistical-likelihood Exo-Planetary Habitability Index (SEPHI) is presented. It has been developed to cover the current and future features required for a classification scheme disentangling whether any discovered exo\-planet is potentially habitable compared with life on Earth. The SEPHI uses likelihood functions to estimate the habitability potential. It is defined as the geometric mean of four sub-indexes related with four comparison criteria: Is the planet telluric?; Does it have an atmosphere dense enough and a gravity compatible with life?; Does it have liquid water on its surface?; Does it have a magnetic field shielding its surface from harmful radiation and stellar winds?. Only with seven physical characteristics, can the SEPHI be estimated: Planetary mass, radius, and orbital period; stellar mass, radius, and effective temperature; planetary system age. We have applied the SEPHI to all the planets in the Exoplanet Encyclopaedia using a Monte Carlo Method. Kepler-1229 b, Kepler-186 f, and  Kepler-442 b have the largest SEPHI values assuming certain physical descriptions. Kepler-1229 b is the most unexpected planet in this privileged position since no previous study pointed to this planet as a potentially interesting and habitable one. In addition, most of the tidally locked Earth-like planets present a weak magnetic field, incompatible with habitability potential. We must stress that our results are linked to the physics used in this study. Any change in the physics used only implies an updating of the likelihood functions. We have developed a web application allowing the on-line estimation of the SEPHI: http://sephi.azurewebsites.net/
\end{abstract}




\section{Introduction}\label{intro}

Searching for life out of our Solar System involves the achievement of several goals. One of these goals is to discover as many potentially habitable exo-Earths as possible \citep{phl}. That is, planets having the same necessary physical conditions for life as Earth has. This catalogue of potentially habitable exoplanets will be the stepping stone for the next stages in the search for life.

Past and current ground-based and space telescopes have yielded, so far, the discovery of thousands of exoplanets \citep{exop.eu}. As soon as the first small planets close to their stellar habitable zone were discovered, the necessity of an index comparing them with Earth arose \citep{Schulze-Makuch}. In above-mentioned work, the authors presented two indexes: 1) The Earth Similarity Index (ESI, equation \ref{esi}), physically comparing the exo-planet with The Earth, and 2) The Pla\-netary Habitability Index (PHI, equation \ref{phi}) accounting for the habitability potential of the exo-planet, compared with what we know about life. Therefore, the goal of these indexes is to define a classification scheme for ascertaining whether exo-planets are potentially habitable compared with Earth's main physical characteristics, from an astrobiological standpoint.

The ESI was initially defined as the product of the re\-lative differences exoplanet - Earth using four physical quantities. Therefore, the ESI has values ranging from 0 to 1, with 0 representing a planet completely different from the Earth and 1 being the Earth itself. Each comparison is weighted by means of a fixed exponent (see \citet{Schulze-Makuch} for details on how these coefficients are obtained). The original definition of the ESI was

\begin{equation}
ESI = \prod^{n}_{i=1}\Big(1-\Big|\frac{x_i - x_{i \oplus}}{x_i + x_{i \oplus}}\Big| \Big)^{\frac{\omega_i}{n}}
\label{esi}
\end{equation}

\noindent where $x_i$ and $x_{i \oplus}$ are the exoplanet and Earth ``$i'th$'' variable to be compared, respectively; $\omega_i$ is the weight associated to variable '$i$'; and $n$ is the number of variables used for obtaining the index, in this case, 4. These four characteristics and their corresponding $\omega_i$ \citep{Schulze-Makuch} are summarized in Table \ref{esi_peso}.

\begin{table}
	\centering
	\caption{\label{esi_peso}Physical characteristics used to obtain the ESI and their corresponding exponents $\omega_i$.}
	\begin{tabular}{l|r}	
		\hline
		Planet's characteristic & $\omega_i$ \\\hline
		Radius & 0.57 \\
		Mean density & 1.07 \\
		Escape velocity & 0.70 \\
		Surface temperature & 5.58 \\
	\end{tabular}
\end{table}

Moreover, the PHI was defined as the geome\-tric mean of some values related to the appearance and safe evolution of life. This index was initially defined as

\begin{equation}
PHI = \Big( S\cdot E\cdot C\cdot L \Big)^{\frac{1}{4}}
\label{phi}
\end{equation}

\noindent where $S$ accounts for the presence of a stable substrate, $E$ for the available energy, $C$ for the appropriate chemistry, and $L$ for the presence of a liquid solvent. Each value is, at the same time, subdivided into different measurements such as the presence of an atmosphere, a magnetosphere, quantity of light and heat received, etc. The value obtained is finally normalized to the largest one \citep{Schulze-Makuch}. Therefore, PHI ranges between 0 (absence of habitability potential) and 1 (potentially habitable).

\citet{Irwin2014} proposed an evolution of their PHI, completing a set of three indexes for analysing exo-planets from an astrobiological point of view. The Biological Complexity Index (BCI) is ``designed to provide a quantitative estimate of the relative probability that complex, macro-organismic life forms could have emerged on other worlds''. In addition: ``The BCI differs from the PHI, in that it estimates a subset of those worlds on which any form of life might appear. The BCI differs from the ESI by ranking planets on their habitability for complex biology, rather than their geophysical similarity to Earth.'' \citep{Irwin2014}. They use the same basic structure of the PHI chan\-ging/adding some ingredients, such as geophysics and planet age. Each value is also subdivided in different measurements such as planetary mean density and orbital eccentricity.

These indexes are definitively a breakthrough in the direction of a homogeneous classification of potentially ha\-bitable exoplanets. Unfortunately, they have some shortcomings that must be addressed:

\begin{itemize}
	\item Related with the ESI:
	\begin{itemize}
		\item The weights assigned to each variable, rather than to equalize them, are used to prioritize them in the overall index. This results in an extremely sensitive index to the surface temperature, being quite insensitive to the other variables. 
		\item The mean density is not conclusive. For example, given the Mars' mean density, a planet can be telluric (as Mars), an ocean or even an ice giant \citep{Dumus_2014}. This is also applicable to the BCI.
	\end{itemize}
	\item Related with the PHI and BCI:
	\begin{itemize}
		\item The evaluation of the value of every measurement is done through an ad-hoc quantized classification.
		\item The PHI mix complementary criteria. For example, the energy coming from the hosting star and that from tidal flexing are both taken into account simultaneously for $E$, when they are not necessarily at the same time.
		\item The value of each mean parameter is obtained as the modulus of the vector of sub-parameters. Therefore, the number of sub-parameters has an impact on the weight of these sub-parameters in the overall index. This weight has no physical justification.
	\end{itemize}
	\item All: They have no physical meaning.
\end{itemize}

In general, the PHI drives to reasonable estimations, but the described weaknesses can result in a value for Jupiter or Saturn for 0.399 and a value for Mars of 0.560. That is, it estimates that the gas giant planets of the Solar System are not much different from Mars from a potentially habitable point of view. In addition, the outer planets Uranus, Neptune and the dwarf planets Pluto and Ceres have a non-negligible PHI value larger than 0.2.

Recently, \citet{Bora2016} developed a more elaborate procedure to estimate these indexes based on data mining techniques, not changing the physical basis of the indexes.

\citet{McLaughlin} proposed an alternative ESI. He focused on the monetary evaluation of exoplanets, from a co\-lo\-nization point of view, to rank their habitability potential. His proposal adds some interesting new concepts, compared with the ESI of \citet{Schulze-Makuch}, that can be used to go forward in this classification scheme:

\begin{itemize}
	\item It has two base factors biasing the index in favour of old, low mass and close stars (this last value measures somehow the scientific study and co\-lonization easiness).
	\item The comparison with Earth's characteristics is done u\-sing Gaussian distributions centred at Earth's values. He compares three characteristics: Logarithm of the relative mass, planet's effective temperature, and planetary system age.
\end{itemize}

\citet{Barnes2015} presented a ``Habitability Index for Transiting Exoplanet'' (HITE) to encapsulate the likelihood of a transiting planet to be potentially habitable u\-sing only observables coming from transits and scaling laws. Although the HITE is a likelihood statistics, it presents some weakness that also deserves to be addressed:

\begin{itemize}
	\item The probability of a planet being telluric is assigned using only the planet radius.
	\item It analyses three parameters: if a planet is rocky, if it has an atmosphere dense enough and if it has liquid water on its surface, ignoring the impact of harmful radiation and stellar wind on the appearance and evolution of life.
	\item The likelihood functions are step functions, something not fully accurate in macro and unbounded physical systems.
\end{itemize}

Since orbital eccentricity is not always available, \citet{Barnes2015} defines a simplified index H' without this parameter. In any case, the HITE (or the H') is a step forward in the search for a reliable classification scheme of planetary habitability potential.

\subsection{The PHI in the near future}

Future space missions devoted to exoplanetary science, such as {\it PLATO2.0} \citep{Plato}, {\it TESS} \citep{TESS} and {\it CHEOPS} \citep{CHEOPS}, together with {\it GAIA} \citep{Gaia}, will provide very accurate data of thousands or even millions of stars and planetary systems.

This suggests that it is highly recommended to have a PHI with the following features:

\begin{itemize}
	\item {\it Automated statistical characterization of exoplanets}. This characterization must be of easy implementation in general pipelines and user-friendly.
	\item {\it Efficient estimation of the PHI}. Given two planets, the difference between their PHIs must be correlated with their real habitability potential difference.
	\item {\it Identification of very interesting exoplanets to prioritize observations}.	
\end{itemize}

The previous PHIs and related indexes fulfil some of these characteristics, but not all at the same time. Some modifications are needed to offer the scientific community a more useful index. In this work, we use statistical likelihood to define a new PHI (SEPHI) fulfilling the above features. As input we use some of the most common exoplanetary observables, which are also part of the official {\it PLATO2.0} science data products. It has been applied to all the known exoplanets. As a result of this, Kepler-1229 b, Kepler-186 f, and  Kepler-442 b arise as very interesting targets for further analysis. Kepler-1229 b is the most unexpected planet in this privileged position since no previous study pointed to this planet as an interesting potentially habitable one. In Section 2, the characteristics of this SEPHI are described. In Section 3, the main results of the application of this SEPHI and its comparison with the results using the original PHI are shown. Section 4 is devoted to summary and conclusions.

\section{Statistical-likelihood Exo-Planetary Habitability Index (SEPHI)}

As a consequence of the success and weaknesses of the PHIs found in the literature, described in the previous section, and the features we expect for a useful PHI in the following years, we have proposed a new classification strategy. It is done using a new index with the following characteristics:

\begin{itemize}
	\item We use likelihood functions instead of ad-hoc quantized values to describe the habitability potential of the exo-planet.
	\item We use comparison criteria instead of single variables. A comparison criterion can be a single variable or a combination of them defining a likelihood function.
	\item We avoid any free parameter or non-physical weight in the evaluation of the overall index. Therefore, all the comparison criteria have the same weight in the final result.
\end{itemize}

To determine the comparison criteria, we focused on those describing the basic conditions allowing life on Earth, that is:

\begin{enumerate}
	\item To be a {\it telluric} planet, with part of its surface composed of solid silicates.
	\item To be able to {\it retain an atmosphere} dense enough and a {\it gravity} compatible with life.
	\item To have {\it liquid water} on its surface.
	\item To have a {\it magnetic field} shielding its surface from stellar winds and cosmic radiations.
\end{enumerate}

In addition, for the likelihood functions, we use Gaussian-like profiles, since they usually provide better des\-criptions of unobstructed physical and biological processes \citep{Banks, Gron}.

With this in mind, we have defined four sub-indexes ($\mathcal{L}_i$) building the SEPHI. The final result will be the geo\-metric mean of these four sub-indexes and represents the statistical likelihood of a planet to be potentially habitable from an astrobiological point of view:

\begin{equation}
SEPHI=\prod^{n}_{i=1}\mathcal{L}_i^{1/n}
\label{slesi}
\end{equation}

This definition of the SEPHI is a statistic only if the $\mathcal{L}_i's$ are statistically independent. The study of the impact of the cross-correlations will be faced in the near future.

\subsection{$\mathcal{L}_1$: Telluric planet}

One of the basic requirements for life to appear on Earth is to have a solid base providing stability for the development of cells. That means that the planet must be telluric with a certain chemical composition. \citet{ZS} deve\-loped a grid of solid planet models using different chemical mixing with three elements: $Fe$, $MgSiO_3$ and $H_2O$. In that work, they show a planetary mass - radius relation for different compositions. Their results are compatible with the Earth's general composition, modelled as 17$\%$ $Fe$ and 83$\%$ $MgSiO_3$ using a fully differentiated two-component model \citep[see][]{Dressing}.

Using this grid, we can estimate the likelihood of a planet to have a telluric composition for a given mass and radius. Therefore we have infinite likelihood functions ($\mathcal{L}_{1,m_p}$), one per relative planetary mass ($m_p\equiv M_p / M_\oplus$). For a given planetary mass we define that a planet has $\mathcal{L}_{1,m_p}(r_p)=1$ when its relative radius ($r_p\equiv R_p/R_\oplus$) is lower than that corresponding to a planet with a composition of 100$\%$ of $MgSiO_3$. From this limit to larger densities the planet is regarded as telluric.

On the other hand, we define $\mathcal{L}_{1,m_p}(r_p)=0$ when its relative radius is larger than that corresponding to a composition of 50$\%$ of $H_2O$ and 50$\%$ of $MgSiO_3$. From this limit to lower densities we estimate that the planet wouldn't have any solid part on its surface.

The decay of the likelihood function from $\mathcal{L}_{1,m_p}(r_p)=1$ to $\mathcal{L}_{1,m_p}(r_p)=0$ is described using a Gaussian-like function centred at a relative radius equal to the larger limit of $\mathcal{L}_{1,m_p}(r_p)=1$ (that with a chemical composition of 100$\%$ of $MgSiO_3$) and a standard deviation $\sigma_1$ equal to one-third of the difference of relative radius limiting $\mathcal{L}_{1,m_p}(r_p)=0$ and $\mathcal{L}_{1,m_p}(r_p)=1$ respectively. Thus, for a given $m_p$ we have:

\begin{eqnarray}
\mu_{1,m_p} = r_{p,100\%\,MgSiO_3} \nonumber \\
\mu_{2,m_p} = r_{p,[50\%\,MgSiO_3-50\%\,H_2O]} \nonumber \\
\sigma_{1,m_p} = \frac{\mu_{2,m_p}-\mu_{1,m_p}}{3} \nonumber
\end{eqnarray}

Therefore, the likelihood function for a given planet mass is:

\begin{eqnarray*}
\mathcal{L}_{1,m_p}(r_p) =1\;\;\;\;\; &\mbox{for}\,\,r_p\le \mu_{1,m_p} \nonumber \\
\mathcal{L}_{1,m_p}(r_p) =e^{-\frac{1}{2}\big(\frac{r_p-\mu_{1,m_p}}{\sigma_{1,m_p}}\big)^2}\;\;\;\;\; &\mbox{for}\,\, \mu_{1,m_p}<r_p<\mu_{2,m_p} \\
\mathcal{L}_{1,m_p}(r_p) =0\;\;\;\;\; &\mbox{for}\,\,\mu_{2,m_p}\le r_p \nonumber
\end{eqnarray*}

\begin{figure}
	\centering
	\includegraphics[width=\columnwidth]{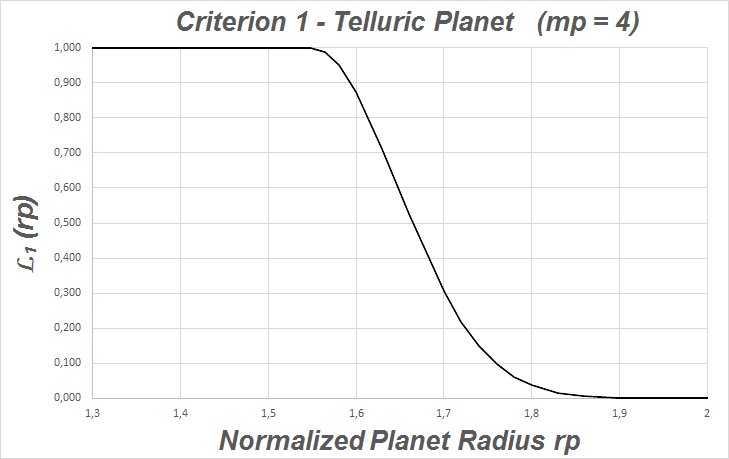}
	\caption{\label{telurico}Telluric likelihood function of a planet with $M_p=4~M_{\oplus}$.}
\end{figure}

In Fig. \ref{telurico} we show the likelihood function of a planet being telluric with $M_p=4~M_{\oplus}$.

\subsection{$\mathcal{L}_2$: Atmosphere and planet gravity}

The second ingredient of the SEPHI is the existence of a dense atmosphere and the presence of a planetary gravity compatible with life.

As a measurement of the planet's capability for retaining an atmosphere, we have used the escape velocity relative to the Earth ($v_e$). This escape velocity is a function of the relative planet's gravity ($g$) and radius ($r_p$):

\begin{equation}
v_e=\sqrt{g\, r_p}
\end{equation}

The surface temperature of the planet also contributes to the instability of its atmosphere. \citet{Kuchner} studied the atmospheric volatile elements in the case of Earth-like planets in the habitable zone. One conclusion of his work is that, in this case, the surface temperature is not critical to explain the absence or not of an atmosphere on the planet. Therefore, as a first approximation, we will not take into account the surface temperature to evaluate the likelihood of a planet having a dense atmosphere similar to Earth.

To estimate the likelihood function $\mathcal{L}_2$ we must first find its limits. To do this, we follow very conservative arguments since the processes of how an atmosphere evaporates due to a lack of gravity or how life is affected by large gravity environments are still inaccurately known.

On the one hand, we estimate that a planet without gravity (a limit theoretical case) has a null likelihood of potential habitability ($\mathcal{L}_2(0) = 0$).

On the other hand, the maximum limit radius for a Super-Earth can be defined at 3$R_\oplus$ \citep{Deming}. Since planets with a composition denser than the ``collisional strip'' may not exist \citep{Marcus}, we adopt a density with a composition 100$\%$ Fe as a conservative reference to estimate the maximum value for the gravity being compa\-tible with life. Using the grid of \citet{ZS} the estimated maximum relative gravity for this type of planets is 25. This drives to a top limit escape velocity of 8.66.

Therefore, the key values of the likelihood function for the escape velocity are:

\begin{itemize}
	\item A planet with $v_e=1$ has a likelihood $\mathcal{L}_2(v_e) = 1$.
	\item A planet with $v_e=0$ has no atmosphere and, therefore, $\mathcal{L}_2(v_e) = 0$.
	\item A planet with $v_e=8.66$ has a gravity not compatible with life as we know it, and $\mathcal{L}_2(v_e) = 0$.
\end{itemize}

To determine the likelihood function among these points, we use Gaussian-like profiles. The first zone, from $v_e=0$ to $v_e=1$ has a $\sigma_{21}=\frac{1-0}{3}$ and the second one, from $v_e=1$ to $v_e=8.66$ has a $\sigma_{22}=\frac{8.66-1}{3}$, corresponding to one-third of the difference between the maximum value of the likelihood and the corresponding limits for $\mathcal{L}_2(v_e) = 0$.

The likelihood function $\mathcal{L}_2$ is then defined as:

\begin{eqnarray}
\mathcal{L}_2(v_e) =e^{-\frac{1}{2}\big(\frac{v_e-1}{\sigma_{21}}\big)^2}\;\;\;\;\; &\mbox{for}\,\,v_e<1 \\
\mathcal{L}_2(v_e) =e^{-\frac{1}{2}\big(\frac{v_e-1}{\sigma_{22}}\big)^2}\;\;\;\;\; &\mbox{for}\,\,v_e\ge 1  \nonumber
\end{eqnarray}

\begin{figure}
	\centering
	\includegraphics[width=\columnwidth]{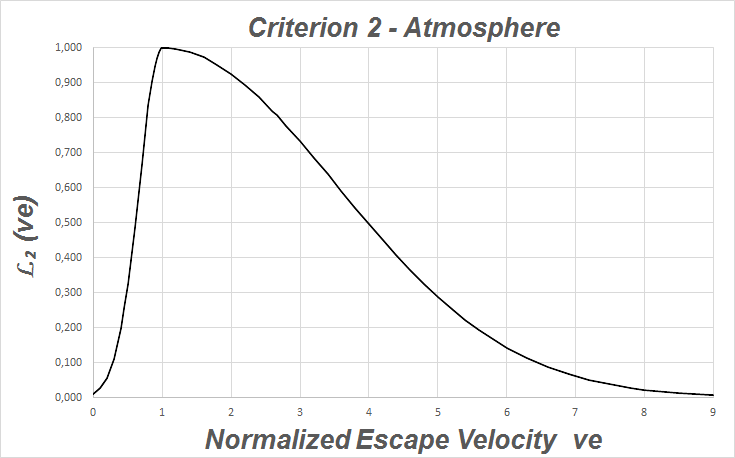}
	\caption{\label{velocidad_escape}Likelihood function of a planet having a escape velocity compatible with the existence of a dense atmosphere and the development of life.}
\end{figure}

In Fig. \ref{velocidad_escape} we show the likelihood function of a planet having an atmospheric density similar to the Earth and a gravity compatible with life, with the assumptions described in this section.

\subsection{$\mathcal{L}_3$: Surface liquid water}

The third element of the SEPHI is the likelihood of a planet having liquid water on its surface. It is directly linked with the so-called Habitability Zone (HZ). In the literature, there are numerous studies related with this HZ. Here we will use the works of \citet{Kopparapu2014, Kopparapu2013} and \citet{Selsis}.

\citet{Kopparapu2014, Kopparapu2013} defined different boun\-daries for delimiting the HZ. We are interested in the following four boundaries:

\begin{itemize}
	\item {\it Recent Venus}: Empirical inner boundary of the HZ, ``based on the inference that Venus has not had liquid water on its surface for at least the past 1 billion years.'' \citep{Kopparapu2013,Solomon1991}. At the Solar System this limit is located at 0.75 AU.
	\item {\it Runaway Greenhouse}: Inner limit at which oceans eva\-porate completely (we call this limit $D_2$). At the Solar System this limit is located at 0.95 AU.
	\item {\it Maximum Greenhouse}: Where Rayleigh scattering by $CO_2$ begins to outweigh the greenhouse effect (we call this limit $D_3$). At the Solar System this limit is located at 1.68 AU.
	\item {\it Early Mars}: A more optimistic empirical limit, compared with the Maximum Greenhouse, based on the observation that early Mars was warm enough for liquid water to flow on its surface \citep{Kopparapu2013}. At the Solar System this limit is located at 1.77 AU.
\end{itemize}


These limits depend on the stellar effective temperature. We use the most updated equations of \citet{Kopparapu2014, Kopparapu2013} to determine how they change in the case of $F$, $G$, $K$ and $M$ stars. They also determine the change of the HZ boundaries with the mass of the planet. They found that the boundaries of the HZ follow the scaling relation:

\begin{eqnarray}
\nonumber S_{eff}=S_{eff,\odot}+a\cdot (T_{eff}-5780)+b\cdot (T_{eff}-5780)^2\\
+c\cdot (T_{eff}-5780)^3+d\cdot (T_{eff}-5780)^4\\
\;\;\;\mbox{and}\;\;\; D=\sqrt{\frac{L_n}{S_{eff}}}\;\;AU
\label{scal_HZ}
\end{eqnarray}

\noindent where $S_{eff}$ is the effective stellar flux \citep[see][for its definition]{Kopparapu2013}, $S_{eff,\odot}$ is the equivalent for the Solar-system, $a$, $b$, $c$ and $d$ are coefficients having different values depending on the HZ boundary and the planetary mass, $T_{eff}$ is the stellar effective temperature, $L_n$ is the stellar luminosity in solar units, and $D$ is the corresponding HZ distance. These equations are valid for scaling the HZ in stars with $T_{eff}$ between 2600$^\circ$K and 7200$^\circ$K.

We need to define the inner and outer points of 0$\%$ potential habitability likelihood. The ``Recent Venus'' and the ``Early Mars'' limits cannot be these points, since these planets were potentially habitable in the past. We will use those described by \citet{Selsis} as the outer limit for water being completely evaporated in the context of an atmosphere with 100$\%$ of vapour water clouds, and the inner limit for water being solid in the context of an atmosphere with 100$\%$ of $CO_2$ clouds. They will be our inner limit $D_1$ and outer limit $D_4$ respectively. In the Solar System, $D_1$ is located at 0.51 AU and $D_4$ at 2.40 AU \citep{Selsis}.

In our study, $D_1$ evolves with the stellar temperature and planetary mass as the ``Recent Venus'' boundary of \citet{Kopparapu2014, Kopparapu2013}, $D_2$ as the ``Runaway Greenhouse'' boundary, $D_3$ as the ``Maximum Greenhouse'' boundary and $D_4$ as the ``Early Mars'' boundary. These equations provide an estimation of the boundaries close to those of \citet{Selsis}. The differences are always lower than 5$\%$ except for $D_4$ in the case of cool stars ($M$ and $K$), where the differences can reach 20$\%$.

Therefore, for every star within this temperature range, five zones we can define for the likelihood function as a function of the orbital semi-mayor axis $a$:

\begin{itemize}
	\item {\it Hot Zone} ($a<D_1$). Water likely in gas form. $\mathcal{L}_3(a) = 0$.
	\item {\it Inner Transition Zone} (ITZ, $D_1\le a<D_2$). The likelihood of a planet having liquid water is larger than 0.
	\item {\it Green Zone} ($D_2\le a \le D_3$). Water likely in liquid form. $\mathcal{L}_3(a) = 1$
	\item {\it Outer Transition Zone} (OTZ, $D_3< a \le D_4$). The likelihood of a planet having liquid water is larger than 0.
	\item {\it Cold Zone} ($a>D_4$). Water likely in solid form. $\mathcal{L}_3(a) = 0$
\end{itemize}

\begin{figure}
	\centering
	\includegraphics[width=\columnwidth]{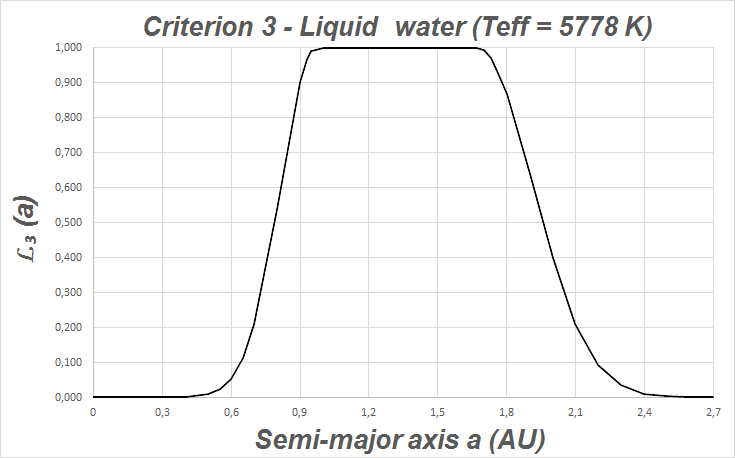}
	\caption{\label{agua}Likelihood function of a planet having liquid water on its surface orbiting the Sun.}
\end{figure}

For the transition zones, Gaussian-like profiles are used to describe the likelihood function. The Gaussian para\-meters are defined as:

\begin{eqnarray*}
\mbox{For the ITZ:}\;\;\;\mu_{31}=D_2\;\;\;\mbox{and}\;\;\;\sigma_{31}=\frac{D_2-D_1}{3} \nonumber \\
\mbox{For the OTZ:}\;\;\;\mu_{32}=D_3\;\;\;\mbox{and}\;\;\;\sigma_{32}=\frac{D_4-D_3}{3}\nonumber
\end{eqnarray*}

With this, we define a likelihood function $\mathcal{L}_3(a)$ as:

\begin{eqnarray}
\mathcal{L}_3(a)=e^{-\frac{1}{2}\big(\frac{a-\mu_{31}}{\sigma_{31}}\big)^2}\;\;\;\;\; &\mbox{for}\,\,a < D_2\nonumber\\
\mathcal{L}_3(a)=1\;\;\;\;\; &\mbox{for}\,\,D_2 \le a \le D_3\\
\mathcal{L}_3(a)=e^{-\frac{1}{2}\big(\frac{a-\mu{32}}{\sigma_{32}}\big)^2}\;\;\;\;\; &\mbox{for}\,\,a > D_3\nonumber
\end{eqnarray}

In Fig. \ref{agua} we show the likelihood function of a planet orbiting the Sun having liquid water on its surface, with the assumptions described in this section.

\subsection{$\mathcal{L}_4$: Magnetic field}

The presence of a magnetic field is intimately linked to life on Earth, protecting it by deflecting harmful radiation and stellar winds. Thus, the presence of a magnetic field should be a must for habitability. This magnetic field is generated by a dynamo effect when convective fluids are moving in the interior of the planet and at least one transition radiation - convection zone is present. In the case of telluric planets, this fluid is liquid iron at the inner convective zone.

\begin{table}
	\centering
	\caption{\label{resumen_m}Estimations coming from the different formulations used in this work (Sano, OC and SOC) compared with the observed (Obs) values for the Solar system planets}
	\begin{tabular}{l|r|r|r|r}
		\hline
		Planet & $\mathcal{M}_n$(Obs) & $\mathcal{M}_n$(Sano) & $\mathcal{M}_n$(OC) & $\mathcal{M}_n$(SOC)\\\hline
		Mercury & 0.0004 & 0.0006 & 0.0013  & 0.0014\\
		Venus & 0 & 0.003 & 0 & 0.041\\
		Earth & 1 & 1 & 0.95 & 1\\
		Mars & 0.15 & 0.09 & 0.12 & 0.10\\
		Jupiter & 18000 & 5184 & 20513 & 19157\\
		Saturn & 580 & 1866 & 1410 & 887\\
		Uranus & 50 & 84 & 218 & 29\\
		Neptune & 24 & 92 & 179 & 26\\
	\end{tabular}
\end{table}

\begin{table}
	\centering
	\caption{\label{valores_eq_simp}Variables to calculate the simplified normalized magnetic moment using equation \ref{olson_simp} as a function of $\beta_1$ and $\beta_2$.}
	\begin{tabular}{l|l|l|l|l}
		\hline
		Type & Reference & $\rho_{0n}$ & $r_{0n}$ & $F_n$ \\\hline
		Telluric & Earth & 1 & $\beta_1$& $\beta_1$\\
		Ice planet & Between Telluric&&&\\
		&and Ice giant & 0.45 & 1.8$\beta_1$ & 4$\beta_1$\\
		Ice giant & Neptune & 0.18 & 4.5$\beta_1$& 20$\beta_1$\\
		Gas giant & Jupiter & 0.16 & 16$\beta_1$$\beta_2$& 100$\beta_1$$\beta_2$\\
	\end{tabular}
\end{table}

\citet{Sano} estimated that the magnetic moment of a planet ($\mathcal{M}$) has the form:

\begin{equation}
\mathcal{M}=\rho^{1/2}R_p^{7/2}\Omega
\label{sano}
\end{equation}

\noindent where $\rho$ is the planet density and $\Omega$ its angular frequency. Testing this model with the Solar system planets we find that it provides reasonable results for telluric planets but it fails when estimating the magnetic moment of giant ones. The more massive the planet, the larger the error (see Table \ref{resumen_m}).

\citet{Olson_christ} derived an alternative expression for determining the magnetic moment:

\begin{equation}
\mathcal{M}=k_1\rho_0^{1/2}r_0^{3}F^{1/3}d^{1/3}
\label{olson}
\end{equation}

\noindent where $k_1$ is a constant, $\rho_0$ the density of the convective zone, $r_0$ the nucleus radius, $F$ the average convective buo\-yancy flux, and $d$ the nucleus convective thickness. Their estimations for Solar system planets are closer to observations than those coming from equation \ref{sano} (see Table \ref{resumen_m}). Unfortunately, for extrasolar planets, some of the variables in equation \ref{olson} are unknown.

\citet{Mercedes} estimated $r_0$, $\rho_0$ and $d$ as a function of planetary mass and radius using different core - mantle chemical compositions. With this, they estimate the planet's magnetic moment for different masses, radius and angular frequencies. Their result is very interesting for obtaining likelihood functions, but the dependence with the gene\-rally unknown angular frequency (except for tidally locked planets) is currently stopper for its inclusion in the SEPHI.

\citet{Zuluaga2013} developed a complete set of equations for the estimation of the planet's magnetic moment including thermal evolution, providing accurate results.

The main goal of this work is to present the SEPHI and its potential. Therefore, in general we have used the simplest possible formulation offering a large accuracy. In the case of the estimation of the planet's magnetic moment, we have used the formulations of \citet{Sano} and \citet{Olson_christ}. In addition, we have done some assumptions for simplifying equation \ref{olson} in order to allow its application to every exoplanet with an observed radius and/or mean density.

First, we will work with magnetic moments norma\-lized to Earth value ($\mathcal{M}_n \equiv \mathcal{M}/\mathcal{M}_\oplus$). In this context, we assume that the normalized radius of the planet nucleus ($r_{0n}$) is equal to the normalized nucleus convective thickness ($d_n$). The estimation of $r_0$ proposed by \citet{Mercedes} is scheduled as future work. With this first assumption, equation \ref{olson} is simplified to

\begin{equation}
\mathcal{M}_n=\alpha\rho_{0n}^{1/2}r_{0n}^{10/3}F_n^{1/3}
\label{olson_simp}
\end{equation}

\noindent where $\alpha$ is a correction of the normalized magnetic moment. If the regime is multipolar, $\alpha = 0.05$. If the planet has a internally heated dynamo, $\alpha = 0.15$. For the rest of the cases, $\alpha = 1$ \citep{Olson_christ}.

A second important assumption is: To obtain the normalized radius and the normalized average convective buo\-yancy flux, two correction factors will be used ($\beta_1$ and $\beta_2$). These factors are related with the sizes and density va\-ria\-tions of the planet with respect those of some reference planets, that is,

\begin{equation}
\beta_1 = \frac{R_p}{R_r}\;\;\;\;\;\mbox{and}\;\;\;\;\;\beta_2 = \frac{\rho_p}{\rho_r}
\label{s_facts}
\end{equation}

\noindent where $X_p$ is the radius or density of the planet and $X_r$ those of the reference planet (Earth, Jupiter, or Neptune). With this second assumption, we can estimate the norma\-lized magnetic moment of a planet as an extrapolation of the Earth, Jupiter or Neptune one, depending on the planet's internal structure.

In Table \ref{valores_eq_simp}, the values of the different ingredients of equation \ref{olson_simp}, as a function of the symplifying factors (equation \ref{s_facts}), are presented for the different planetary general structures.

In Table \ref{resumen_m} we show a summary of the estimations of the normalized magnetic moment for the Solar System planets, obtained with the three formulations used in this work: Sano, Olson - Christensen (OC) and Simplified Olson - Christensen (SOC, this work). In this table, we observe that:

\begin{itemize}
	\item For Mercury, the only tidally locked planet in the Solar system, Sano provides the best estimation.
	\item For the reference planets (Earth, Jupiter, and Neptune), the SOC provides estimations with an accuracy better than 10$\%$.
	\item For the rest of the planets, SOC provides estimations up to seven times more accurate than OC. This accuracy is enough for the purposes of this work.
\end{itemize}

An important conclusion from the work of \citet{Olson_christ} is that, when a planet is not tidally locked to the star, its magnetic moment is almost dipolar and independent of the rotational angular velocity, avoi\-ding the necessity of the observation of the planet's angular frequency, something not available in ge\-ne\-ral. This, together with the conclusions listed above, leads to the use of Sano's equation (eq. \ref{sano}) for tidally locked or potentially locked planets in multipolar regime, and the SOC equation (eqs. \ref{olson_simp} and \ref{s_facts}, and Table \ref{valores_eq_simp}) for the rest.

To determine whether a planet is tidally locked or not, we follow the work of \citet{Griess}, which allows an estimation of the outer limit of the tidally locked zone as a function of the stellar mass and the density of the planet. \citet{Griess} defined a transition zone where a planet can be potentially locked. The age of the system is the key to disentangle whether a potentially locked planet is in fact locked or not.

\begin{figure}
	\centering
	\includegraphics[width=\columnwidth]{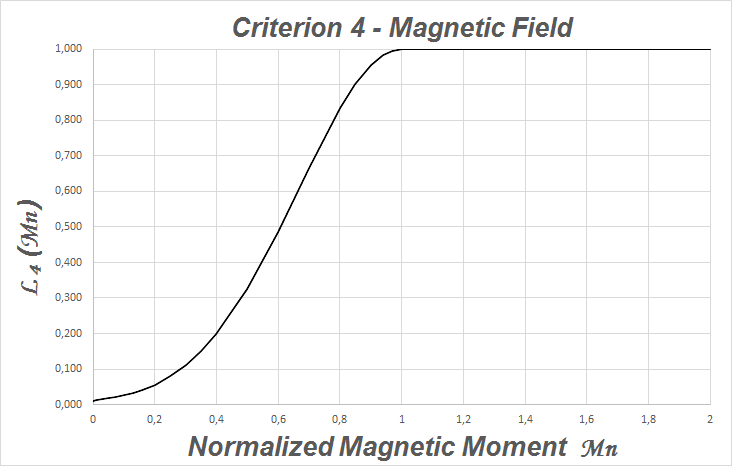}
	\caption{\label{momento_magnetico}Likelihood function of a planet having a magnetic moment protecting life as Earth does.}
\end{figure}

For the likelihood function, we define that planets bigger than Earth and tidally unlocked should readily have no problems having a strong dipolar magnetic field protecting its possible life, except if an unusual event has happened in its past. On the other hand, low-mass planets (with a small nucleus) and/or slow rotating planets (mainly tidally locked) would have weak multipolar magnetic moments.

This leads to two well-differentiated zones in the likelihood function:

\begin{itemize}
	\item We know that the Earth's magnetic moment has allowed the inception and evolution of life. Therefore, planets with a magnetic moment larger than Earth would be able to efficiently protect their possible hosted life. That is: $\mathcal{L}(\mathcal{M}_n \ge 1) = 1$
	\item A planet with a null magnetic moment will have a null likelihood, that is, $\mathcal{L}(0) = 0$ 
\end{itemize}

The transition between these zones will be described using a Gaussian-like profile, with mean $\mu_4=1$ and standard deviation one-third of this transition region $\sigma_4=\frac{1-0}{3}=0.333$.

That is, the likelihood function of criterion 4 is

\begin{eqnarray}
\mathcal{L}(\mathcal{M}_n) = e^{-\frac{1}{2}\big(\frac{\mathcal{M}_n-\mu_4}{\sigma_4}\big)^2}\;\;\;\;\; &\mbox{for}\,\,\mathcal{M}_n < 1 \\
\mathcal{L}(\mathcal{M}_n) = 1\;\;\;\;\; &\mbox{for}\,\,\mathcal{M}_n \ge 1 \nonumber
\end{eqnarray}

In Fig. \ref{momento_magnetico} we show the likelihood function of a planet having a magnetic moment compatible with life, with the assumptions described in this Section.

\section{Results}

We have applied the SEPHI to all the exoplanets in \citet{exop.eu} and the most relevant object of the Solar System. To estimate its value and uncertainty, we perform a Monte Carlo method (up to 10000 realizations per estimation) using the input uncertainties. We have assumed that the published uncertainties correspond to a $3\sigma$ departure from the central value using a Gaussian probability distribution.

\begin{table}
	\centering
	\caption{\label{resultados_sol} Summary of the main results obtained for objects of the Solar System. PHI is the value coming from \citet{Schulze-Makuch}; $\mathcal{L}_i$ is the SEPHI sub-index; SEPHI is the result obtained using equation \ref{slesi}.}
	\begin{tabular}{l|r|r|r|r|r|r}
		\hline
		Planet & PHI & $\mathcal{L}_1$ & $\mathcal{L}_2$ & $\mathcal{L}_3$ & $\mathcal{L}_4$ & SEPHI\\
		\hline
		Earth & 0.97 & 1 & 1 & 1 & 1 & 1\\
		Mars & 0.56 & 0.978 & 0.263 & 1 & 0.026 & 0.285\\
		Venus & 0.39 & 1 & 0.976 & 0.334 & 0.011 & 0.249\\
		Moon & 0.00 & 0.703 & 0.060 & 1 & 0.012 & 0.149\\
		Mercury & 0.00 & 1 & 0.177 & 0.008 & 0.011 & 0.064\\
		Ceres & 0.25 & 1 & 0.018 & 0 & 0.011 & 0\\
		Eris & - & 1 & 0.031 & 0 & 0.011 & 0\\
		Pluto & 0.20 & 1 & 0.028 & 0 & 0.011 & 0\\
		Haumea & - & 1 & 0.023 & 0 & 0.011 & 0\\
		Makemake & - & 1 & 0.022 & 0 & 0.011 & 0\\
		Uranus & 0.28 & 0 & 0.937 & 0 & 1 & 0\\
		Neptune & 0.28 & 0 & 0.909 & 0 & 1 & 0\\
		Saturn & 0.40 & 0 & 0.679 & 0 & 1 & 0\\
		Jupiter & 0.40 & 0 & 0.224 & 0 & 1 & 0\\
	\end{tabular}
\end{table}

As we have described in the previous section, to obtain the SEPHI we need to know seven physical parameters: Planetary mass, radius, and orbital period; stellar mass, radius, and effective temperature; planetary system age. For our analysis we have neglected the following cases:

\begin{itemize}
	\item Planets with no Mass and radius informed simulta\-neous\-ly.
	\item Hosting stars with no mass and radius informed simultaneously.
	\item Planets with no orbital period and major semi-axis informed simultaneously.
	\item Planets with $M_p > 4200 M_\oplus$ or $R_p>15R_\oplus$.
	\item Hosting stars with $T_{eff}<2200^\circ$K or $T_{eff}> 8000^\circ$K.
\end{itemize}

\begin{table*}
	\centering
	\caption{\label{resultados_exo} Summary of the main results obtained. PHI is the value coming from \citet{Schulze-Makuch}; H' the index coming from \citet{Barnes2015}; $\mathcal{L}_i$ is the SEPHI sub-index; SEPHI is the result obtained using equation \ref{slesi}; Uncertainty is the 3$\sigma$ error obtained using a Monte-Carlo estimations.}
	\begin{tabular}{l|r|r|r|r|r|r|r}
		\hline
		Planet & PHI & $\mathcal{L}_1$ & $\mathcal{L}_2$ & $\mathcal{L}_3$ & $\mathcal{L}_4$ & SEPHI & Uncertainty ($\pm$)\\
		\hline
		Kepler-186 f & 0.79 & 0.941 & 0.999 & 0.977 & 0.996 & 0.98 & 0.02\\
		Kepler-1229 b & 0.83 & 0.864 & 0.990 & 1 & 1 & 0.96 & 0.03\\
		Kepler-442 b & 0.90 & 0.791 & 0.992 & 1 & 1 & 0.94 & 0.06\\
		Kepler-62 e & 0.51 & 0.908 & 0.647 & 0.978 & 1 & 0.87 & 0.11\\
		Kepler-62 f & 0.40 & 0.943 & 0.613 & 1 & 0.988 & 0.87 & 0.14\\
		Kepler-22 b & 0.66 & 1 & 0.472 & 1 & 1 & 0.83 & 0.02\\
		Kepler-441 b & 0.68 & 0.447 & 0.985 & 0.991 & 1 & 0.81 & 0.06\\
		Kepler-452 b & 0.89 & 0.446 & 0.986 & 0.954 & 1 & 0.81 & 0.04\\
		KOI-4427.01 & 0.68 & 0.207 & 0.981 & 0.966 & 1 & 0.7 & 0.3\\
		KIC-5522786 b & 0.82 & 0.694 & 0.991 & 0.287 & 0.890 & 0.65 & 0.10\\
		Kepler-69 c & 0.79 & 0.291 & 0.985 & 0.537 & 1 & 0.63 & 0.08\\
		Kepler-283 c & 0.85 & 0.110 & 0.982 & 0.995 & 1 & 0.57 & 0.12\\
		Kepler-1544 b & 0.86 & 0.087 & 0.983 & 1 & 1 & 0.54 & 0.03\\
		K2-9 b & 0.89 & 0.492 & 0.985 & 0.796 & 0.083 & 0.42 & 0.02\\
		Pr\'oxima Centauri b & 0.91 & 0.963 & 0.999 & 1 & 0.030 & 0.411 & 0.003\\
		TRAPPIST 1 c & 0.82 & 1 & 0.998 & 0.066 & 0.387 & 0.400 & 0.007\\
		Kepler-445 d & 0.75 & 1 & 0.963 & 0.160 & 0.156 & 0.39 & 0.04\\
		TRAPPIST 1 g & 0.76 & 0.699 & 0.941 & 1 & 0.030 & 0.38 & 0.04\\
	\end{tabular}
\end{table*}

Unfortunately, not all the remaining exoplanets feature all the above-mentioned required physical parameters. Therefore, we have used the following approximations for estimating the unknown parameters:

\begin{itemize}
	\item When the planetary mass or radius is unknown, we use the \citet{Chen} estimations for planets to guess the uninformed value.
	\item When the stellar mass or radius is unknown, we use the \citet{Chen} estimations for low-mass stars to guess the uninformed value.
	\item When the stellar temperature is unknown, we use its mass to estimate the value.
	\item When the system age is unknown, we use 2 Gy as a reference value.
\end{itemize}

In Table \ref{resultados_sol} the main results of our study applied to the most relevant objects of the Solar System are shown compared with the PHI value \citep{Schulze-Makuch}. Most of the objects have a SEPHI = 0 since the only energy source included is the hosting star. Therefore, all objects out of the Solar HZ have a null likelihood assuming the physics included in our work. Additionally, it seems unrealistic that Saturn has a PHI = 0.4 whereas Mars has a PHI = 0.56. The SEPHI corrects this, presenting Mars as a planet with a small but not null habitability potential, mainly due to its absence of a strong enough magnetic field.

Table \ref{resultados_exo} shows the main results of our study applied to some relevant exoplanets, from the point of view of their habitabi\-lity potential. In the case of an input value with an unknown uncertainty, we have fixed it as a 10$\%$ for mass and radius and $\pm100^\circ$K for the stellar temperature.These results are compared with the PHI value.

This table reveals that Kepler-1229 b, Kepler-186 f, and  Kepler-442 b are those with a SEPHI close to one. Some planets with a small PHI rise-up to a SEPHI around 0.87 (such as Kepler-62 e and f, and Kepler 22 b). The other way around, some planets signed by the PHI as interesting targets are discarded by our SEPHI (such as Proxima Centaury b, TRAPPIST 1 c and g, K2-9 b, and Kepler-445 d). The main reason for these small similarities is the absence of a magnetic field. Most of the tidally locked planets have a small probability of having a magnetic field protecting their surfaces from harmful radiations and stellar winds. The position of TRAPPIST 1 c and  Kepler-455 d close to the boundaries of the HZ additionally decrease their habitability potential.

We must stress at this point that our results depend on the physics implemented to obtain the likelihood functions. This is one of the main benefits of the SEPHI, the physics for estimating the habitability potential is included using the likelihood functions, so that the algorithm for obtaining SEPHI remains unchanged. Therefore, any new or alternative physics for evaluating the habitability potential can be easi\-ly implemented just defining a new corresponding likelihood function, allowing the comparison and evaluation of the impact of the different physics. For example, the inclusion of the magnetic field in the index is a step forward in the evaluation of the potential habitability. With the SEPHI we can focus on the different approximations to evaluate these magnetic fields.

Another important index, described in section \ref{intro}, is the HITE index \citep{Barnes2015}. This index has some features in common with the SEPHI and they provide an ha\-bit\-ability potential likelihood as output. Therefore they can be directly compared. In table \ref{resultados_h} we show the H' and SEPHI values obtained for a sample of Kepler exoplanets, since H' is focused on transiting exoplanets.

In this table we see:

\begin{itemize}
	\item H' uses only the planet radius to estimate whether it is telluric or not, and we use its radius and mass through the Zeng-Sasselov grid. Therefore, we expect quite different results for the estimation of $\mathcal{L}_1$.
	\item H' doesn't include the magnetic field, something SEPHI does. Therefore, those planets with an expected small magnetic field must have very different values of H' and SEPHI. This is the case of KOI-2626.01, a tidally locked exoplanet.
	\item $\mathcal{L}_2$ and $\mathcal{L}_3$ are studied in a similar way in both indexes. Therefore, those planets with a high likelihood to be telluric and protected by a magnetic field have similar values of H' and SEPHI.
	\item The uncertainty of the input values is taken into account in the SEPHI. Therefore, the Monte Carlo simulation can lead to a different result compared with the H' for the same inputs as a consequence of these uncertainties. This is the case, for example, of KOI-5737.01.
\end{itemize}

Therefore, H' and SEPHI both point in the same direction, offering reasonably similar results, but we estimate that SEPHI is a step forward since it adds some important features not included in the HITE index offering a richer ana\-lysis of the potential habitability of an exoplanet.

\begin{table}
	\centering
	\caption{\label{resultados_h}Summary of the comparison between the H' index \citep{Barnes2015} and the SEPHI.}
	\begin{tabular}{l|r|r}
		\hline
		Planet & H' & SEPHI\\
		\hline
		KOI-3456.02 & 0.955 & 0.835 \\
		KOI-7235.01 & 0.932 & 0.843 \\
		KOI-5737.01 & 0.916 & 0.757 \\
		KOI-2194.03 & 0.894 & 0.933 \\
		KOI-2626.01 & 0.887 & 0.346 \\
		KOI-6108.01 & 0.865 & 0.827 \\
		KOI-5948.01 & 0.843 & 0.789 \\
		KOI-6425.01 & 0.839 & 0.586 \\
		KOI-5554.01 & 0.217 & 0.601 \\
		KOI-7587.01 & 0.200 & 0.061 \\
	\end{tabular}
\end{table}

\section{Summary and Conclusions}

In this work, we have presented a new Statistical-likelihood Exo-Planetary Habitability Index (SEPHI) based on the likelihood estimation of four different comparison criteria with Earth as the only place we know harbouring life. This SEPHI solves the shortcomings of the previous habitability indexes in the literature, allowing the statistical interpretation of the results, avoiding the free parameters, allowing the comparison of groups of physical parameters, and separating the physics used to estimate the habitability potential and the index structure. The physics impacts on the final result through the likelihood functions.

The proposed SEPHI is obtained by evaluating the similarity likelihood of four criteria: To be a Telluric planet, to have a dense atmosphere and a gravity compatible with bio\-logical processes, to have liquid water on its surface, and to have a magnetic field protecting the planet surface from harmful radiations and stellar winds. The total SEPHI is the geometric mean of these four sub-indexes. To obtain this index we use seven physical characteristics: Planetary mass, radius, and orbital period; stellar mass, radius, and effective temperature; planetary system age. These parameters are those more commonly known for exoplanets and are part of the official {\it PLATO2.0} science data products.

Our procedure enables the easy inclusion of e\-very new discovery since the statistical likelihood is obtained using likelihood functions calculated based on the current knowledge. All new knowledge will impact on it. The up-dating of the SEPHI is reached only changing the corres\-ponding function, without any additional change in the rest of the procedure. Therefore, the problem of comparing with Earth is now focused on our physical knowledge of the processes impacting the planet's habitability. The uncertainty of the input observables is included in the SEPHI using a Monte Carlo method to estimate the index and its uncertainty.

We have applied this SEPHI to all the currently known exoplanets. Kepler-1229 b, Kepler-186 f, and  Kepler-442 b are those with a SEPHI closer to one. Kepler-1229 b is the most unexpected planet in this privileged position since no previous study pointed to this planet as an interesting potentially habitable one. The Solar System planets have a low habitability potential, in the case of Mars and Venus mainly due to their weak magnetic field, and Mars also due to its low-density atmosphere. In general, most of the tidally locked Earth-like planets have a weak magnetic field, incompatible with habitability potential. This is a problem for most of the planets discovered orbiting M-dwarf stars with masses lower than $2M_\oplus$ \citep{Mercedes}. In any case, we must stress that our results are linked to the physics used in this study.

We have developed a web application for allowing the community the easy online estimation of the SEPHI only informing the seven required physical parameters \citep{slesi_web}.

\section*{Acknowledgement}

AM acknowledges funding support from Spanish public funds for research under project ESP2015-65712-C5-5-R (MINECO/FEDER), and from project RYC-2012-09913 under the 'Ram\'on y Cajal' program of the Spanish MINECO. The authors want to acknowledge the effort and very constructive comments and suggestions of the anonymous referee. We also acknowledge Brian Boland and Jose Delgado for the English edition of this manuscript.




\bibliographystyle{mnras}
\bibliography{esi_bib} 

\begin{thebibliography}{}
\makeatletter
\relax
\def\mn@urlcharsother{\let\do\@makeother \do\$\do\&\do\#\do\^\do\_\do\%\do\~}
\def\mn@doi{\begingroup\mn@urlcharsother \@ifnextchar [ {\mn@doi@}
  {\mn@doi@[]}}
\def\mn@doi@[#1]#2{\def\@tempa{#1}\ifx\@tempa\@empty \href
  {http://dx.doi.org/#2} {doi:#2}\else \href {http://dx.doi.org/#2} {#1}\fi
  \endgroup}
\def\mn@eprint#1#2{\mn@eprint@#1:#2::\@nil}
\def\mn@eprint@arXiv#1{\href {http://arxiv.org/abs/#1} {{\tt arXiv:#1}}}
\def\mn@eprint@dblp#1{\href {http://dblp.uni-trier.de/rec/bibtex/#1.xml}
  {dblp:#1}}
\def\mn@eprint@#1:#2:#3:#4\@nil{\def\@tempa {#1}\def\@tempb {#2}\def\@tempc
  {#3}\ifx \@tempc \@empty \let \@tempc \@tempb \let \@tempb \@tempa \fi \ifx
  \@tempb \@empty \def\@tempb {arXiv}\fi \@ifundefined
  {mn@eprint@\@tempb}{\@tempb:\@tempc}{\expandafter \expandafter \csname
  mn@eprint@\@tempb\endcsname \expandafter{\@tempc}}}

\bibitem[\protect\citeauthoryear{Banks \& Tran}{Banks \& Tran}{2009}]{Banks}
Banks H.,  Tran H.,  2009, Mathematical and Experimental Modeling of Physical
  and Biological Processes.
\url {https://books.google.com/books?hl=en{\&}lr={\&}id=SSRapIe8p3QC{\&}pgis=1}

\bibitem[\protect\citeauthoryear{Barnes, Meadows  \& Evans}{Barnes
  et~al.}{2015}]{Barnes2015}
Barnes R.,  Meadows V.~S.,   Evans N.,  2015, The Astrophysical Journal, 814,
  91

\bibitem[\protect\citeauthoryear{Bora, Saha, Agrawal, Safonova, Routh  \&
  Narasimhamurthy}{Bora et~al.}{2016}]{Bora2016}
Bora K.,  Saha S.,  Agrawal S.,  Safonova M.,  Routh S.,   Narasimhamurthy A.,
  2016, \mn@doi [Astronomy and Computing] {10.1016/j.ascom.2016.08.001}, 17,
  129

\bibitem[\protect\citeauthoryear{{Broeg} et~al.,}{{Broeg}
  et~al.}{2013}]{CHEOPS}
{Broeg} C.,  et~al., 2013, in European Physical Journal Web of Conferences. p.
  03005 (\mn@eprint {arXiv} {1305.2270}), \mn@doi{10.1051/epjconf/20134703005}

\bibitem[\protect\citeauthoryear{{Chen} \& {Kipping}}{{Chen} \&
  {Kipping}}{2017}]{Chen}
{Chen} J.,  {Kipping} D.,  2017, \mn@doi [ApJ] {10.3847/1538-4357/834/1/17},
  \href {http://adsabs.harvard.edu/abs/2017ApJ...834...17C} {834, 17}

\bibitem[\protect\citeauthoryear{{Deming} et~al.,}{{Deming}
  et~al.}{2009}]{Deming}
{Deming} D.,  et~al., 2009, \mn@doi [PASP] {10.1086/605913}, \href
  {http://adsabs.harvard.edu/abs/2009PASP..121..952D} {121, 952}

\bibitem[\protect\citeauthoryear{{Dressing} et~al.,}{{Dressing}
  et~al.}{2015}]{Dressing}
{Dressing} C.~D.,  et~al., 2015, \mn@doi [ApJ] {10.1088/0004-637X/800/2/135},
  \href {http://adsabs.harvard.edu/abs/2015ApJ...800..135D} {800, 135}

\bibitem[\protect\citeauthoryear{{Dumusque} et~al.,}{{Dumusque}
  et~al.}{2014}]{Dumus_2014}
{Dumusque} X.,  et~al., 2014, \mn@doi [ApJ] {10.1088/0004-637X/789/2/154},
  \href {http://adsabs.harvard.edu/abs/2014ApJ...789..154D} {789, 154}

\bibitem[\protect\citeauthoryear{Exoplanets.eu}{Exoplanets.eu}{2017}]{exop.eu}
Exoplanets.eu 2017, \url {http://www.exoplanet.eu/}

\bibitem[\protect\citeauthoryear{{Grie{\ss}meier}, {Stadelmann}, {Grenfell},
  {Lammer}  \& {Motschmann}}{{Grie{\ss}meier} et~al.}{2009}]{Griess}
{Grie{\ss}meier} J.-M.,  {Stadelmann} A.,  {Grenfell} J.~L.,  {Lammer} H.,
  {Motschmann} U.,  2009, \mn@doi [Icarus] {10.1016/j.icarus.2008.09.015},
  \href {http://adsabs.harvard.edu/abs/2009Icar..199..526G} {199, 526}

\bibitem[\protect\citeauthoryear{Gr{\"{o}}nholm \& Annila}{Gr{\"{o}}nholm \&
  Annila}{2007}]{Gron}
Gr{\"{o}}nholm T.,  Annila A.,  2007, \mn@doi [Mathematical Biosciences]
  {10.1016/j.mbs.2007.07.004}, 210, 659

\bibitem[\protect\citeauthoryear{Irwin, M{\'{e}}ndez, Fair{\'{e}}n  \&
  Schulze-Makuch}{Irwin et~al.}{2014}]{Irwin2014}
Irwin L.,  M{\'{e}}ndez A.,  Fair{\'{e}}n A.,   Schulze-Makuch D.,  2014,
  \mn@doi [Challenges] {10.3390/challe5010159}, 5, 159

\bibitem[\protect\citeauthoryear{Jordan}{Jordan}{2008}]{Gaia}
Jordan S.,  2008, \mn@doi [AN] {10.1002/asna.200811065}, 329, 875

\bibitem[\protect\citeauthoryear{Kopparapu et~al.,}{Kopparapu
  et~al.}{2013}]{Kopparapu2013}
Kopparapu R.~K.,  et~al., 2013, \mn@doi [The Astrophysical Journal]
  {10.1088/0004-637X/765/2/131}, 131

\bibitem[\protect\citeauthoryear{Kopparapu, Ramirez, Schottelkotte, Kasting,
  Domagal-goldman  \& Eymet}{Kopparapu et~al.}{2014}]{Kopparapu2014}
Kopparapu R.~K.,  Ramirez R.~M.,  Schottelkotte J.,  Kasting J.~F.,
  Domagal-goldman S.,   Eymet V.,  2014, \mn@doi [The Astrophysical Journal
  Letters] {10.1088/2041-8205/787/2/L29}, 29, 0

\bibitem[\protect\citeauthoryear{{Kuchner}}{{Kuchner}}{2003}]{Kuchner}
{Kuchner} M.~J.,  2003, \mn@doi [ApJ] {10.1086/378397}, \href
  {http://adsabs.harvard.edu/abs/2003ApJ...596L.105K} {596, L105}

\bibitem[\protect\citeauthoryear{{L{\'o}pez-Morales}, {G{\'o}mez-P{\'e}rez}  \&
  {Ruedas}}{{L{\'o}pez-Morales} et~al.}{2011}]{Mercedes}
{L{\'o}pez-Morales} M.,  {G{\'o}mez-P{\'e}rez} N.,   {Ruedas} T.,  2011,
  \mn@doi [Origins of Life and Evolution of the Biosphere]
  {10.1007/s11084-012-9263-8}, \href
  {http://adsabs.harvard.edu/abs/2011OLEB...41..533L} {41, 533}

\bibitem[\protect\citeauthoryear{{Marcus}, {Stewart}, {Sasselov}  \&
  {Hernquist}}{{Marcus} et~al.}{2009}]{Marcus}
{Marcus} R.~A.,  {Stewart} S.~T.,  {Sasselov} D.,   {Hernquist} L.,  2009,
  \mn@doi [ApJ] {10.1088/0004-637X/700/2/L118}, \href
  {http://adsabs.harvard.edu/abs/2009ApJ...700L.118M} {700, L118}

\bibitem[\protect\citeauthoryear{McLaughlin}{McLaughlin}{2012}]{McLaughlin}
McLaughlin G.,  2012, \url {http://oklo.org/2012/02}

\bibitem[\protect\citeauthoryear{{Olson} \& {Christensen}}{{Olson} \&
  {Christensen}}{2006}]{Olson_christ}
{Olson} P.,  {Christensen} U.~R.,  2006, \mn@doi [Earth and Planetary Science]
  {10.1016/j.epsl.2006.08.008}, \href
  {http://adsabs.harvard.edu/abs/2006E%26PSL.250..561O} {250, L561}

\bibitem[\protect\citeauthoryear{PHL}{PHL}{2017}]{phl}
PHL 2017, \url {http://phl.upr.edu/home}

\bibitem[\protect\citeauthoryear{{Rauer} et~al.,}{{Rauer} et~al.}{2014}]{Plato}
{Rauer} H.,  et~al., 2014, \mn@doi [Experimental Astronomy]
  {10.1007/s10686-014-9383-4}, \href
  {http://adsabs.harvard.edu/abs/2014ExA....38..249R} {38, 249}

\bibitem[\protect\citeauthoryear{Ricker et~al.,}{Ricker et~al.}{2014}]{TESS}
Ricker G.~R.,  et~al., 2014, \mn@doi [arXiv] {10.1117/12.2063489}, 9143, 914320

\bibitem[\protect\citeauthoryear{Rodr\'iguez-Mozos}{Rodr\'iguez-Mozos}{2017}]{slesi_web}
Rodr\'iguez-Mozos J.~M.,  2017, \url {http://sephi.azurewebsites.net/}

\bibitem[\protect\citeauthoryear{Sano}{Sano}{1993}]{Sano}
Sano Y.,  1993, \mn@doi [J. Geomag. Geoelec.] {doi.org/10.5636/jgg.45.65}, 45,
  65

\bibitem[\protect\citeauthoryear{Schulze-Makuch et~al.,}{Schulze-Makuch
  et~al.}{2011}]{Schulze-Makuch}
Schulze-Makuch D.,  et~al., 2011, \mn@doi [Astrobiology]
  {10.1089/ast.2010.0592}, 11, 1041

\bibitem[\protect\citeauthoryear{{Selsis}, {Kasting}, {Levrard}, {Paillet},
  {Ribas}  \& {Delfosse}}{{Selsis} et~al.}{2007}]{Selsis}
{Selsis} F.,  {Kasting} J.~F.,  {Levrard} B.,  {Paillet} J.,  {Ribas} I.,
  {Delfosse} X.,  2007, \mn@doi [A\,\&\,A] {10.1051/0004-6361:20078091}, \href
  {http://adsabs.harvard.edu/abs/2007A%26A...476.1373S} {476, 1373}

\bibitem[\protect\citeauthoryear{{Solomon} \& {Head}}{{Solomon} \&
  {Head}}{1991}]{Solomon1991}
{Solomon} S.~C.,  {Head} J.~W.,  1991, \mn@doi [Science]
  {10.1126/science.252.5003.252}, 252, 252

\bibitem[\protect\citeauthoryear{{Zeng} \& {Sasselov}}{{Zeng} \&
  {Sasselov}}{2013}]{ZS}
{Zeng} L.,  {Sasselov} D.,  2013, \mn@doi [PASP] {10.1086/669163}, \href
  {http://adsabs.harvard.edu/abs/2013PASP..125..227Z} {125, 227}

\bibitem[\protect\citeauthoryear{{Zuluaga}, {Bustamante}, {Cuartas}  \&
  {Hoyos}}{{Zuluaga} et~al.}{2013}]{Zuluaga2013}
{Zuluaga} J.~I.,  {Bustamante} S.,  {Cuartas} P.~A.,   {Hoyos} J.~H.,  2013,
  \mn@doi [The Astrophysical Journal] {10.1088/0004-637X/770/1/23}, 770, 23

\makeatother
\end{thebibliography}

\bsp	
\label{lastpage}
\end{document}